\documentclass{PoS}
\usepackage{amsfonts, amsthm, amsmath}
\usepackage[utf8]{inputenc}
\usepackage{slashed}
\usepackage{mathrsfs}
\usepackage{amssymb}
\usepackage{color}
\usepackage{cite}

\newcommand{\eq}{\begin{equation}}
\newcommand{\feq}{\end{equation}}
\newcommand{\eqn}{\begin{eqnarray}}
\newcommand{\feqn}{\end{eqnarray}}

\newcommand{\be}{\begin{equation}}
\newcommand{\ee}{\end{equation}}
\newcommand{\ben}{\begin{displaymath}}
\newcommand{\een}{\end{displaymath}}
\newcommand{\bea}{\begin{eqnarray}}
\newcommand{\eea}{\end{eqnarray}}

\newcommand{\bean}{\begin{eqnarray*}}
\newcommand{\eean}{\end{eqnarray*}}

\newcommand{\ma}[1]{\mbox{$\mathcal{#1}$}}

\newcommand{\mrm}[1]{\mbox{$\mathrm{#1}$}}

\title{$\textbf{AdS}_3$ vacua and surface defects in massive IIA}

\ShortTitle{$\textbf{AdS}_3$ vacua and surface defects in massive IIA}

   \author{Giuseppe Dibitetto$^a$\, \speaker{Nicolò Petri}$^{\,\,b}$\vspace{2mm} \\
    $\mbox{}^a$Institutionen f\"or fysik och astronomi, University of Uppsala,  Box 803, SE-751 08 Uppsala, Sweden.\\\vspace{2mm}
   $\mbox{}^b$Department of Mathematics, Bo\u{g}azi\c{c}i University,
34342, Bebek, Istanbul, Turkey.\\
        E-mail: \email{giuseppe.dibitetto@physics.uu.se, nicolo.petri@boun.edu.tr}}

\abstract{We summarize the results and the ideas in \cite{Dibitetto:2017tve,Dibitetto:2017klx,Dibitetto:2018iar} where new warped $\mrm{AdS}_3$ backgrounds are derived in massive IIA string theory by uplifting exact solutions in $\ma N=1$, $d=7$ and $\ma N=(1,1)$, $d=6$ gauged supergravities. These solutions are respectively asymptotically $\mrm{AdS}_7$ and $\mrm{AdS}_6$ and they are related to the D2-D4-NS5-D6-D8 brane intersection. We provide a particular supergravity solution in 10d describing this bound state and we discuss the relations between its near-horizon geometry and the uplifts from 6d and 7d. Then we give the holographic interpretations of these $\mrm{AdS}_3$ warped backgrounds in terms of $\ma N=(0,4)$ defect $\mrm{SCFT}_2$ within the $\ma N=(1,0)$ $\mrm{SCFT}_6$ and the $\ma N=2$ $\mrm{SCFT_5}$.  }

\FullConference{Corfu Summer Institute 2018 "School and Workshops on Elementary Particle Physics and Gravity"\\
		(CORFU2018)\\
		31 August - 28 September, 2018\\
		Corfu, Greece}

\begin{document}

\section{Introduction}

In this proceeding we will consider two particular realizations of defect conformal field theories realized in massive IIA string theory \cite{Dibitetto:2017tve,Dibitetto:2017klx,Dibitetto:2018iar}.

A defect CFT is defined by degrees of freedom localized on a boundary embedded in the background of a higher-dimensional CFT \cite{Cardy:1984bb}. From the point of view of this higher-dimensional theory, the defect is described by a deformation related to a position-dependent coupling. This deformation breaks partially the conformal isometries in the bulk and, as a consequence, the 1-point functions are no longer vanishing. Moreover a non-trivial displacement operator appears and this is due to the fact of that the energy-momentum tensor is not conserved anymore.

The first examples\footnote{For a non-exhaustive list of references on conformal defects in string theory and holography see \cite{DeWolfe:2001pq,Bachas:2001vj,Aharony:2003qf,Clark:2004sb,Kapustin:2005py,Clark:2005te,DHoker:2006qeo,DHoker:2006vfr,Buchbinder:2007ar,DHoker:2007zhm,Lunin:2007ab,Gaiotto:2008sa,Gaiotto:2008sd,Aharony:2011yc,Gutperle:2012hy,deLeeuw:2015hxa,Billo:2016cpy,DelZotto:2018tcj,Karndumri:2018yiz,Dibitetto:2018gtk}.} of defect in string theory were given in \cite{Karch:2000gx}. The idea is to look at the emerging of defect CFTs as the result of the interesection of some ``defect branes'' ending on a given brane system, with an AdS vacuum at the near-horizon. The defect branes break partially the isometries of the AdS vacuum of the original brane setup producing a lower-dimensional warped AdS background. The degrees of freedom of the defect CFT are related to the boundary conditions of the intersection and the warping of the corresponding supergravity background is associated to the backreaction of the defect branes onto the bulk. This is the supergravity interpretation of the position-dependent deformation of the SCFT, holographically dual to the original higher-dimensional AdS near-horizon.

More explicitly, let's consider a supersymmetric $\mathrm{AdS}_d$ vacuum associated to the near-horizon of a brane system and let's assume the existence of a compactification linking the 10d (or 11d) physics to a solution in a $d$-dimensional gauged supergravity.
The interesection with defect branes ending on the system will be given by a bound state with a $(p+1)$-dimensional worldvolume described by a $d$-dimensional spacetime background of the type
\begin{equation}
 ds_d^2=e^{2U(r)}\,ds^2_{{\scriptsize \mrm{AdS}_{p+2}}}+e^{2V(r)}\,dr^2+e^{2W(r)}\,ds^2_{d-p-3}\,.
 \label{defect} 
\end{equation} 
 This background is characterized by a $\mrm{AdS}_{p+2}$ foliation and an asymptotic behavior locally given by the $\mrm{AdS}_d$ vacuum of the theory. Moreover \eqref{defect} can be uplifted and it reproduces a warped geometry of the type $\mrm{AdS}_{p+2}\times \ma {M}_{d-p-2}\times \Sigma_{D-d}$, where $\ma {M}_{d-p-2}$ is given by a fibration of the $(d-p-3)$-dimensional internal manifold over the interval $I_r$ and $\Sigma_{D-d}$ is the internal manifold characterizing the particular truncation. The holographic picture of this supergravity configuration is that of a defect $\mrm{SCFT}_{p+1}$ within the $\mrm{SCFT}_{d-1}$ dual to $\mrm{AdS}_d$ vacuum.
 
 In this proceeding we summarize the ideas and the main results of \cite{Dibitetto:2017tve,Dibitetto:2017klx,Dibitetto:2018iar} where new warped $\mrm{AdS}_3$ backgrounds are constructed explicitly from $\ma N=1$, 7d and $\ma N=(1,1)$, 6d gauged supergravities. These backgrounds are interpreted in massive IIA string theory as 2d defect superconformal field theories within the $\ma N=(1,0)$ $\mrm{SCFT}_6$ and the $\ma N=2$ $\mrm{SCFT}_5$ respectively dual to the $\mrm{AdS}_7$ and $\mrm{AdS}_6$ vacua describing the asymptotics of our solutions.
In order to construct the holographic interpretation in terms of conformal defects, we will consider two particular brane systems in massive IIA string theory, namely the NS5-D6-D8 interesection \cite{Hanany:1997gh} and the D4-D8 system \cite{Brandhuber:1999np}. The supergravity solutions describing these intersections are described at the horizon respectively by $\mrm{AdS}_7\times S^3$ and $\mrm{AdS}_6\times S^4$ warped gemoetries. We will present a general 10d background reproducing the interesection of these two bound states with some new defect branes, respectively D2-D4 and D2-NS5-D6. The effect of these new branes will be that of break some of the isometries of the $\mrm{AdS}_7$ and $\mrm{AdS}_6$ vacua in a way that the near-horizon will be now given by a background of the type $\mathrm{AdS}_3\times S^3\times S^2\times I^2$. Up to a change of coordinate, this near-horizon geometry reproduces the uplifts of the 6d and 7d solutions with whom we started our analysis. Starting from this we will give the holographic interpretation of these $\mrm{AdS}_3$ slicings found in 6d and 7d supergravities in terms of a defect $\ma N=(0,4)$ $\mrm{SCFT}_2$ within $\ma N=(1,0)$ $\mrm{SCFT}_6$ and $\ma N=2$ $\mrm{SCFT_5}$. Then we will conclude by discussing the derivation of the 1-point functions of these defects by using holographic methods and conformal perturbation expansion.

\section{The Setup}

In this section we review the main properties of the NS5-D6-D8 and D4-D8 brane setups in massive IIA string theory. These brane setups give rise to the warped vacua $\mrm{AdS}_7\times S^3$ and $\mrm{AdS}_6\times S^4$, and to their dual\footnote{For a non-exhaustive list on $\mrm{AdS}_7/\mrm{CFT}_6$ and $\mrm{AdS}_6/\mrm{CFT}_5$ correspondences see \cite{Hanany:1997gh,Brunner:1997gf,Brunner:1997gk,Apruzzi:2013yva,Gaiotto:2014lca,Apruzzi:2014qva,Apruzzi:2015zna,DeWolfe:1999hj,Jafferis:2012iv,Assel:2012nf,Bergman:2012qh,Bergman:2012kr,Passias:2012vp,Lozano:2012au,Bergman:2013koa,Lozano:2013oma,Apruzzi:2014qva,DHoker:2016ujz,Passias:2018swc}.} descriptions respectively given in terms of $\ma N=(1,0)$ $\mrm{SCFT}_6$ and $\ma N=2$ $\mrm{SCFT_5}$. Moreover massive IIA admits two consistent truncations around the abova vacua reproducing respectively $d=7$, $\ma N=1$ and $d=6$, $\ma N=(1,1)$ gauged supergravities. These two supergravities will constitute the playground used to generate warped $\mrm{AdS}_3$ solutions describing conformal surface defects in the brane setups considered.


Let's firstly consider the construction originally proposed in \cite{Hanany:1997gh}. This brane system involves D6 branes stretched along D8 branes with NS5 branes completely inside the D6s. The NS5 describe the interface between the D6 and the D8 and their mutual distance defines the gauge coupling of the 6d worlvolume theory of the brane intersection. 
\begin{table}[h!]
\renewcommand{\arraystretch}{1}
\begin{center}
\scalebox{1}[1]{
\begin{tabular}{c||c c c c c c|c||c c c}
branes & $t$ & $y^{1}$ & $y^{2}$ & $y^{3}$ & $y^{4}$ & $y^{5}$ & $z$ & $r$ & $\theta^{1}$ & $\theta^{2}$ \\
\hline \hline
NS5 & $\times$ & $\times$ & $\times$ & $\times$ & $\times$ & $\times$ & $-$ & $-$ & $-$ & $-$ \\
D6 & $\times$ & $\times$ & $\times$ & $\times$ & $\times$ & $\times$ & $\times$ & $-$ & $-$ & $-$ \\
D8 & $\times$ & $\times$ & $\times$ & $\times$ & $\times$ & $\times$ & $-$ & $\times$ & $\times$ & $\times$ \\
\end{tabular}
}
\end{center}
\caption{{\it The brane picture of the $\ma N=(1,0)$ $\mrm{SCFT}_6$ described by a NS5-D6-D8 system. This system is $\mrm{BPS}/4$.
Note that the radial coordinate realizing to the $\mrm{AdS}_{7}$ geometry is given by a combination of $z$ and $r$.}}
\end{table}
This brane setup is described, in the low-energy regime, by an infinite class of warped $\mrm{AdS}_{7}\times S^3$ vacua at the horizon preserving 16 real supercharges and a $SO(3)$ symmetry (as hinted in \cite{Gaiotto:2014lca} and clarified in \cite{Bobev:2016phc}). These string vacua were originally found in
\cite{Apruzzi:2013yva} as BPS solutions of massive type IIA supergravity by using the pure spinor formalism (see also \cite{Apruzzi:2014qva,Apruzzi:2015zna,Passias:2016fkm} for further details).
The internal 3-sphere is squashed and it is described locally as a $S^2$-fibration over a segment. From the point of view of the brane picture the D6-branes fill $\mrm{AdS}_{7}$
and are completely localized at the poles of the 3-sphere. Finally the D8-branes wrap the $S^{2}$.
In \cite{Gaiotto:2014lca} the dual interpretation of the $\mrm{AdS}_7$ vacua in massiva IIA is formulated in terms of linear quivers realizing a $\ma N=(1,0)$ $\mrm{SCFT}_6$ emerging as a fixed point of the 6d Yang-Mills worldvolume theory.

These $\mrm{AdS}_7$ solutions define a warped compactification on the squashed $S^{3}$ \cite{Passias:2016fkm}. This dimensional reduction implies that the physics of massive IIA string theory around the $\mrm{AdS}_{7}\times S^3$ vacua is captured by the minimal\footnote{Only the supergravity multiplet is retained in the compactification.} realization of $\ma N=1$, 7d gauged supergravity. In particular the bosonic fields involved into the 7d supergravity multiplet are
\begin{equation}
 g_{\mu\nu}\,,\qquad X\,,\qquad \ma B_{(3)\,\mu\nu\rho}\,,\qquad A^i_\mu\,,\label{fields7d}
\end{equation}
where $X$ is a real scalar, $\ma B_{(3)}$ is a 3-form gauge potential and $A^i_\mu$ an $SU(2)$-triplet of gauge vectors. The theory is driven by a scalar potential for $X$ depending on the coupling $g$ associated to the gauging of the R-symmetry group $SU(2)_R$ and on a topological mass $h$ for the 3-form. This 7d supergravity realization will constitute the main setup used to find massive IIA solutions describing conformal defects within the $\ma N=(1,0)$ $\mrm{SCFT}_6$.

The other brane setup that we consider is the well-known D4-D8 system where $N$ D4 branes are completely localized on D8 branes with O8 planes on top \cite{Seiberg:1996bd,Brandhuber:1999np}. The supergravity solution describing the low-energy regime of this system is related to an $\mrm{AdS}_6\times S^4$ geometry at the horizon \cite{Brandhuber:1999np,Youm:1999zs,Imamura:2001cr}.
\begin{table}[h!]
\renewcommand{\arraystretch}{1}
\begin{center}
\scalebox{1}[1]{
\begin{tabular}{c||c c c c c || c c||c c c}
branes & $t$ & $y^{1}$ & $y^{2}$ & $y^{3}$ & $y^{4}$ & $z$ & $\rho$ & $\theta^{1}$ & $\theta^{2}$ & $\theta^{3}$ \\
\hline 
D8 & $\times$ & $\times$ & $\times$ & $\times$ & $\times$ & $-$ & $\times$ & $\times$ & $\times$ & $\times$ \\
D4 & $\times$ & $\times$ & $\times$ & $\times$ & $\times$ & $-$ & $-$ & $-$ & $-$ & $-$ 
\end{tabular}
}
\end{center}
\caption{{\it The brane picture underlying the 5d $\ma N=2$ SCFT defined by the D4-D8 system. The system is $\mrm{BPS}/4$ and
the $\textrm{AdS}_{6}\times S^4$ vacuum is realized by a combination of $\rho$ and $z$.}} \label{Table:BO}
\end{table}
The holographic interpretation of this vacuum was constructed in \cite{Seiberg:1996bd,Brandhuber:1999np,Ferrara:1998gv} in terms of a $\ma N=2$ $\mrm{SCFT}_5$ with a $\mrm{Usp}(N)$ gauge group and couplings to fundamental and antysimmetric hypermultiplets.

This vacuum preserves 16 supercharges and $SO(4)$ symmetry. It induces a warped compactification \cite{Cvetic:1999un} of massive IIA on the 4-sphere to the so-called Romans supergravity \cite{Romans:1985tw}, namely the minimal realization of $\ma N=(1,1)$, 6d gauged supergravity. The bosonic content of the supergravity multiplet of this theory is given by 
\begin{equation}
 g_{\mu\nu}\,,\qquad X\,,\qquad \ma B_{(2)\,\mu\nu}\,,\qquad A^0_\mu\,,\qquad A^i_\mu\,,\label{fields6d}
\end{equation}
where $X$ is a real scalar, $\ma B_{(2)}$ is a 2-form gauge potential, $A^0_\mu$ is an abelian vector and $A^i_\mu$ an $SU(2)$-triplet of gauge vectors. The gauging of the theory is very similar to the above 7d case. The theory is driven by a scalar potential for $X$ depending on a coupling $g$ associated to the gauging of the R-symmetry group $SU(2)_R$ and on a topological mass $m$ for the 2-form. This 6d supergravity will constitute the main setup used to find solutions associated to conformal defects within the $\ma N=2$ $\mrm{SCFT}_5$.

\section{Warped $\mathrm{AdS}_3\times S^3\times S^2\times I^2$ Backgrounds}

Let's consider now the two lower-dimensional supergravities introduced in the previous section, namely $\ma N=1$, 7d and $\ma N=(1,1)$, 6d gauged supergravities in their minimal realization. For the bosonic field content given respectively in \eqref{fields6d} and \eqref{fields7d}, we will consider the following backgrounds \cite{Dibitetto:2017tve,Dibitetto:2018iar},
\begin{equation}
  ds_{6, 7}^2= e^{2U(r)}\,ds^2_{{\scriptsize \mrm{AdS}_3}}+e^{2V(r)}\,dr^2+e^{2W(r)}\,ds^2_{\Sigma_{2,3}}\ ,
  \label{AdS33S2ansatz}
\end{equation}
with $\Sigma_{2,3}=\{S^2,S^3\}$ respectively for the 6d and 7d cases. As far it regards the scalar we will suppose in both cases that $X=X(r)$. Furthermore for simplicity we will consider all the vectors vanishing, while for the $p$-form gauge potential we will require that
\begin{equation}
\begin{split}
  &7d\,\mbox{:}\qquad\ma B_{(3)}=k(r)\,\textrm{vol}_{\mathrm{AdS}_{3}} + l(r)\,\textrm{vol}_{{S}^3} \,,\\
   & 6d\,\mbox{:}\qquad\ma B_{(2)} =  b(r)\,\textrm{vol}_{S^2} \,.\label{forms}
  \end{split}
\end{equation}
In \cite{Dibitetto:2017tve,Dibitetto:2018iar}, the BPS equations for \eqref{AdS33S2ansatz} are derived and solved for the cases of two independent warp factors $U$ and $W$, and for the case of one single warp factor $U=W$.

Let consider the second case. The assumption $U=W$ implies that $k=l$ for the 7d 3-form \eqref{forms}. The BPS equations become in this case very simple and their solution in the 7d case is given by
\begin{equation}
\label{7dSOL}
\begin{array}{lclclclclc}
e^{2U} & = & \dfrac{2^{-1/4}}{g^{1/2}}\,\left(\dfrac{r}{1\,-\,r^{5}}\right)^{1/2} & & , & & e^{2V} & = & \dfrac{25}{2\,g^{2}}\,\dfrac{r^{6}}{\left(1\,-\,r^{5}\right)^{2}} & , \\[3mm]
k & = & -\dfrac{2^{1/4}\,L}{g^{3/2}}\left(\dfrac{r^{5}}{1\,-\,r^{5}}\right)^{1/2} & & , & & X & = & r\,, & ,
\end{array}
\end{equation}
where $r$ ranges from $0$ to $1$ and the two gauge coupling $g$ and $h$ satisfy the relation $h=\frac{g}{2\sqrt{2}}$. In the 6d case the solution of the BPS equation takes the following form
\begin{equation}
 \begin{split}
  e^{2U}= &\ \frac{2^{-1/3}}{g^{2/3}}\,\left(\frac{r^\prime}{1-r^{\prime\,5}}\right)^{2/3}\ , \qquad e^{2V}=\frac{8}{g^2}\, 
  \frac{r^{\prime\,4}}{\left(1- r^{\prime\,4}\right)^2}\ ,\\
   b=&\ -\frac{2^{1/3}\,3\,L}{g^{4/3}}\,\left(\frac{r^{\prime\,4}}{1-r^{\prime\,4}}\right)^{1/3}\ ,\qquad \ X=r^\prime\ ,
   \label{CDW}
 \end{split}
\end{equation}
where also in this case $r^\prime$ ranges from $0$ to $1$ and the gauge coupling $g$ and $m$ satisfy the relation $m=\frac{\sqrt 2\, g}{3}$. For $r \,\&\,r^\prime\rightarrow 1$, these solutions are locally described by $\mrm{AdS}_7$ and $\mrm{AdS}_6$ geometries, while, for $r \,\&\,r^\prime\ \rightarrow 0$, they manifest singular behaviors.

The goal is to interpret the singularities appearing in the $r \rightarrow 0$ and $r^\prime \rightarrow 0$ limits in terms of some brane sources. If one consider the uplifts of the asymptotic regime of the solutions \eqref{7dSOL} and \eqref{CDW}, one obtains the $\mrm{AdS}_7\times S^3$ and $\mrm{AdS}_6\times S^4$ vacua, while for other value of $r$ and $r^\prime$ one obtains two $\mathrm{AdS}_3\times S^3\times S^2\times I^2$ warped background that are very similar for their geometric properties. In particular
\begin{equation}
\begin{split}
& 7d\,\mbox{:}\qquad \mathrm{AdS}_3\times S^3\times I_r\qquad  \longrightarrow \qquad 10d\,\mbox{:}\qquad \mathrm{AdS}_3\times S^3\times I_r\times S^2\times I_\xi\,\\
& 6d\,\mbox{:}\qquad \mathrm{AdS}_3\times S^2\times I_{r^\prime} \qquad \longrightarrow \qquad 10d\,\mbox{:}\qquad \mathrm{AdS}_3\times S^2\times I_{r^\prime}\times S^3\times I_{\xi^\prime}\,,\\\label{uplift}
 \end{split}
\end{equation}
where the coordinates $\xi$ and $\xi^\prime$ respectively describes the fibrations coordinates of the warped compactifications of massive IIA on the squashed 3-sphere and 4-sphere. The structure of the fluxes and the same number of supersymmetries preserved hints that the 10d backgrounds obtained by uplifting from 6d and 7d could be the same background up to a change of coordinates, namely $(r,\,\xi)\rightarrow (r^\prime\,,\xi^\prime)$. Unfortunately it is very complicated to derive the explicit form of this change of coordinate because of the highly non-linear dependence of the warp factors in the uplift expressions.

\section{The D2-D4-NS5-D6-D8 System and Conformal Defects}

The two solutions in 6d and 7d written in \eqref{7dSOL} and \eqref{CDW} are asymptotically locally $\mrm{AdS}_7$ and $\mrm{AdS}_6$. As we said in the previous section, from the 10d point of view these are the two warped vacua $\mrm{AdS}_7\times S^3$ and $\mrm{AdS}_6\times S^4$. This means that, if we want to search for a brane setup described by a $\mrm{AdS}_3$ near-horizon of the type of \eqref{uplift}, we have to consider supergravity solutions including as particular limits the solutions of the bound states NS5-D6-D8 and D4-D8 with some new defect branes breaking the isometries of the above vacua.

For the 7d case this solution can be derived exactly and it describes the intersection of a D2-D4 bound state with the NS5-D6-D8 one. The brane solution of the general intersection is given by \cite{Dibitetto:2017klx}
\begin{table}[h!]
\renewcommand{\arraystretch}{1}
\begin{center}
\scalebox{1}[1]{
\begin{tabular}{c||c c|c c c c|c||c c c}
branes & $t$ & $y$ & $\rho$ & $\varphi^{1}$ & $\varphi^{2}$ & $\varphi^{3}$ & $z$ & $r$ & $\theta^{1}$ & $\theta^{2}$ \\
\hline \hline
NS5 & $\times$ & $\times$ & $\times$ & $\times$ & $\times$ & $\times$ & $-$ & $-$ & $-$ & $-$ \\
D6 & $\times$ & $\times$ & $\times$ & $\times$ & $\times$ & $\times$ & $\times$ & $-$ & $-$ & $-$ \\
D8 & $\times$ & $\times$ & $\times$ & $\times$ & $\times$ & $\times$ & $-$ & $\times$ & $\times$ & $\times$ \\
\hline
D2 & $\times$ & $\times$ & $-$ & $-$ & $-$ & $-$ & $\times$ & $-$ & $-$ & $-$ \\
D4 & $\times$ & $\times$ & $-$ & $-$ & $-$ & $-$ & $-$ & $\times$ & $\times$ & $\times$ \\
\end{tabular}
}
\end{center}
\caption{{\it The brane picture of the $\ma N=(0,4)$ defect $\mrm{SCFT}_2$ associated to D2- and D4-branes ending on an NS5-D6-D8 intersection. The above system is $\mrm{BPS}/8$.}}
\end{table}
\be
\label{brane_sol}
\begin{array}{lclc}
ds_{10}^{2} & = & S^{-1/2}H_{\textrm{D}2}^{-1/2}H_{\textrm{D}4}^{-1/2}\,ds_{\textrm{Mkw}_{2}}^{2}\,+\, 
S^{-1/2}H_{\textrm{D}2}^{1/2}H_{\textrm{D}4}^{1/2}\,\left(d\rho^{2}+\rho^{2}\,d\Omega_{(3)}^{2}\right) \,+ & \\[2mm]
& + & K\,S^{-1/2}H_{\textrm{D}2}^{-1/2}H_{\textrm{D}4}^{1/2}\,dz^{2}\,+\,K\,S^{1/2}H_{\textrm{D}2}^{1/2}H_{\textrm{D}4}^{-1/2}\,\left(dr^{2}+r^{2}\,d\Omega_{(2)}^{2}\right) & , \\[3mm]
e^{\Phi} & = & g_{s}\,K^{1/2}\,S^{-3/4}H_{\textrm{D}2}^{1/4}H_{\textrm{D}4}^{-1/4} & , \\[3mm]
H_{(3)} & = & \frac{\partial}{\partial z}\left(KS\right)\textrm{vol}_{(3)}\,-\,dz\,\wedge\,*_{(3)}\left(dK\right) & , \\[3mm]
F_{(0)} & = & m & , \\[3mm]
F_{(2)} & = & -g_{s}^{-1}\,*_{(3)}\left(dS\right) & , \\[3mm]
F_{(4)} & = & g_{s}^{-1}\,\textrm{vol}_{(1,1)}\,\wedge\,dz\,\wedge\,dH_{\textrm{D}2}^{-1} \, + \, 
*_{(10)}\left(\textrm{vol}_{(1,1)}\,\wedge\,\textrm{vol}_{(3)}\,\wedge\,dH_{\textrm{D}4}^{-1}\right) & , 
\end{array}
\ee
where the functions $K(z,r)$ and $S(z,r)$ satsify \cite{Imamura:2001cr}
\be
\left\{
\begin{array}{rccc}
mg_{s}\,K \,-\, \frac{\partial S}{\partial z} & = & 0 & , \\[3mm]
\Delta_{(3)}S \, + \, \frac{1}{2}\frac{\partial^{2}}{\partial z^{2}} S^{2} & = & 0 & ,
\end{array}
\right.
\ee
while 
\be
\begin{array}{lccclc}\label{warpfactors}
H_{\textrm{D}2}(\rho,r) \ = \ \left(1+\frac{q_{\textrm{D}4}}{\rho^{2}}\right)\left(1+\frac{q_{\textrm{D}6}}{r}\right) & , & & & 
H_{\textrm{D}4}(\rho) \ = \ \left(1+\frac{q_{\textrm{D}4}}{\rho^{2}}\right) & .
\end{array}
\ee
As it has been showed in \cite{Dibitetto:2017klx}, the solution \eqref{brane_sol} reproduces the near-horizon geometry $\mathrm{AdS}_3\times S^3\times S^2\times I^2$. Moreover by comparing the fluxes and the supersymmetries preserved the conclusion is that, up to a change of the fibration-coordinates, this near-horizon reproduces the uplift of the 7d solution \eqref{7dSOL}.

The 6d case is analogous, but now the isometries of the $\mrm{AdS}_6$ vacuum are broken by the intersections of the defect branes D2-NS5-D6 \cite{Dibitetto:2018iar}.
\begin{table}[h!]
\renewcommand{\arraystretch}{1}
\begin{center}
\scalebox{1}[1]{
\begin{tabular}{c||c c|c c c c|c||c c c}
branes & $t$ & $y$ & $\rho$ & $\varphi^{1}$ & $\varphi^{2}$ & $\varphi^{3}$ & $z$ & $r$ & $\theta^{1}$ & $\theta^{2}$ \\
\hline \hline
$\mrm{D}8$ & $\times$ & $\times$ & $\times$ & $\times$ & $\times$ & $\times$ & $-$ & $\times$ & $\times$ & $\times$ \\

$\mrm{D}4$ & $\times$ & $\times$ & $-$ & $-$ & $-$ & $-$ & $-$ & $\times$ & $\times$ & $\times$ \\
\hline
$\mrm{D}2$ & $\times$ & $\times$ & $-$ & $-$ & $-$ & $-$ & $\times$ & $-$ & $-$ & $-$ \\

$\mrm{NS}5$ & $\times$ & $\times$ & $\times$ & $\times$ & $\times$ & $\times$ & $-$ & $-$ & $-$ & $-$ \\
$\mrm{D}6$ & $\times$ & $\times$ & $\times$ & $\times$ & $\times$ & $\times$ & $\times$ & $-$ & $-$ & $-$ \\

\end{tabular}
}
\end{center}
\caption{{\it The brane picture of the $\ma N=(0,4)$ defect $\mrm{SCFT}_2$ produced by D2-NS5-D6 branes ending on an D4-D8 bound state. The system is $\mrm{BPS}/8$.}} 
\end{table}
The general form of the 10d solution is the same of \eqref{brane_sol}, but clearly now the explicit form of the warp factors appearing in \eqref{brane_sol} is different, i.e. the parametrization of the $\mrm{AdS}_3$ near-horizon is different respect to that one obtained from the solution \eqref{warpfactors}. Unfortunately we don't have an explicit form of the warp factors in this case, but the structure of fluxes and the supersymmetry preserved by the uplift of the 6d $\mrm{AdS}_3$ solution in \eqref{uplift} are the same of the near-horizon obtained from \eqref{warpfactors}.

The holographic interpretation of these $\mrm{AdS}_3$ solution is given in terms of a $\ma N=(0,4)$ $\mrm{SCFT}_2$ realizing a surface defect respectively within the $\ma N=(1,0)$ $\mrm{SCFT}_6$ and $\ma N=2$ $\mrm{SCFT}_5$. In order to show that the 2d defect $\mrm{SCFT}$s corresponding to the 6d and 7d cases are actually the same, one should provide an explicit change of coordinates within the two uplifts \eqref{uplift}, but the above arguments there is a quite strong evidence on their equivalence.

Finally we checked this holographic interpretation by sketching the calculation of the 1-point functions. The presence of the defect breaks some of the conformal isometries so as a first check it is interesting to see if the position-dependence of the coupling of the deformation describing the defect is the same when it is calculated using the standard holographic dictionary on one side (extracting the 1-point functions from the asymptotic expansion of the supergravity background) and on the other side by considering a conformal perturbation expansion of the correlation functions \cite{Clark:2004sb}. Supposing that the deformation is driven in both cases by the scalar $X$, we obtain the same behaviors in both case 6d and 5d cases, namely \cite{Dibitetto:2017klx,Dibitetto:2018iar}
\begin{equation}
 \begin{split}
  &\langle \ma O_X \rangle_{6d} \sim x^{-4}\,,\\
  &\langle \ma O_X \rangle_{5d} \sim x^{-3}\,,\\
 \end{split}
\end{equation}
where $x$ is the radial coordinate of the $\mrm{AdS}_3$ slicings of \eqref{7dSOL} and \eqref{CDW}.

\section*{Acknowledgements}

NP would like to acknowledge the organizers of the "School and Workshops on Elementary Particle Physics and Gravity" for their kind hospitaliy and for the financial support. The work of NP is supported by the Scientific and Technological Research Council of Turkey (T\" ubitak) Grant No.116F137.
The work of GD is supported by the Swedish Research Council (VR).

 \bibliographystyle{utphys}
  \bibliography{biblio}
  



\end{document}